\pdfoutput=1
\newcommand*{\ATLASLATEXPATH}{latex/}
\documentclass[english,cernpreprint,texlive=2016,txfonts]{latex/atlasdoc}

\usepackage{\ATLASLATEXPATH atlaspackage}
\usepackage{\ATLASLATEXPATH atlasbiblatex}

\usepackage{\ATLASLATEXPATH atlascontribute}

\usepackage{\ATLASLATEXPATH atlasphysics}

\graphicspath{{logos/}{figures/}}
\usepackage{dcolumn}
\usepackage{mathrsfs}

\usepackage{paper-defs}

\AtlasTitle{Search for heavy resonances decaying to a $W$ or $Z$ boson and a Higgs boson in the $q\bar{q}^{(\prime)}b\bar{b}$ final state in $pp$ collisions at $\sqrt{s} = 13$ \tev\ with the ATLAS detector}

\author{The ATLAS Collaboration}

\AtlasNote{ATLAS-EXOT-2016-12}

\PreprintIdNumber{CERN-EP-2017-111}

 \AtlasJournal{Phys.\ Lett.\ B.}

\AtlasAbstract{%
A search for heavy resonances decaying to a $W$ or $Z$ boson and a Higgs boson in the $q\bar{q}^{(\prime)}b\bar{b}$
final state is described.  The search uses \lumi\ \ifb\ of proton--proton collision data at $\sqrt{s} =$ 13 \tev\ 
collected by the ATLAS detector at the CERN Large Hadron Collider in 2015 and 2016.  The data are in agreement 
with the Standard Model expectations, with the largest excess found at a resonance mass of 3.0 \tev\ with a local (global)
significance of \localSig\ (\globalSig)~$\sigma$.
The results are presented in terms of constraints on a simplified model with
a heavy vector triplet.  Upper limits are set on the production cross-section times branching ratio 
 for resonances decaying to a $W$ ($Z$) boson and a Higgs boson, itself decaying to $b\bar{b}$,   in the mass range between 1.1 and 3.8~\tev;
the limits range between \WHlimitOneTev\ and \WHlimitThreePointEightTev~fb (\ZHlimitOneTev\ and \ZHlimitThreePointEightTev~fb) at 95\% confidence level.}

\hypersetup{pdftitle={ATLAS document},pdfauthor={The ATLAS Collaboration}}

\begin{document}

\maketitle



\section{Introduction}
\label{sec:intro}

The discovery of the Higgs boson~\cite{Aad:2012tfa,CMS-HIG-12-028} confirms the validity of the Standard Model (SM) in the description 
of particle interactions at energies up to a few hundred \gev.  However, radiative corrections to the Higgs boson mass drive 
its value to the model's validity limit, indicating either extreme fine-tuning or the presence of new physics at an energy scale 
not far above the Higgs boson mass.  It is natural to expect such new physics to manifest itself through significant coupling
to the Higgs boson, for example in decays of new particles to a Higgs boson and other SM particles.  This Letter presents a 
search for resonances produced in \lumi\ \ifb\ of proton--proton ($pp$) collision data at $\sqrt{s}=13$~\tev\ which decay to a $W$ or $Z$ boson and a Higgs boson.  Such resonances are predicted in multiple models
of physics beyond the SM, e.g.~composite Higgs~\cite{Dugan:1984hq,Agashe:2004rs} or Little Higgs~\cite{Schmaltz:2005ky} models, or models with extra dimensions~\cite{Agashe:2007ki,Agashe:2008jb}.

This search is conducted in the channel where the $W$ or $Z$ and Higgs bosons decay to quarks.  
The high mass region, with resonance masses $m_{VH} >$ 1 \tev\ ($V = W,Z$), where the $V$ and $H$ bosons are highly Lorentz boosted, is considered.  
The $V$ and $H$ boson candidates are each reconstructed in a single jet, using jet substructure techniques and $b$-tagging
to suppress the dominant background from multijet events and enhance the sensitivity to the dominant $H \to b\bar{b}$ decay mode.  The reconstructed dijet mass distribution is used to search for a
signal and, in its absence, to set bounds on the production cross-section times branching ratio for new bosons which decay to a $W$ or $Z$ boson and a Higgs boson.

The results are expressed as limits in a simplified model  which 
incorporates a heavy vector triplet 
(HVT)~\cite{Pappadopulo:2014qza,deBlas:2012qp} of bosons and allows the results to be interpreted in a large class of models.
The new heavy vector bosons couple to the Higgs boson and SM gauge bosons with coupling strength $c_H g_V$ and to the SM fermions with coupling strength $(g^2/g_V )c_F$, where $g$ is the SM SU(2)$_{\textrm L}$ coupling constant. 
The parameter $g_V$ characterizes the interactions of the new vector bosons, while the dimensionless coefficients $c_H$ and $c_F$ parameterize departures of this typical strength for interactions with the Higgs and SM gauge bosons and with fermions, respectively, and are expected to be of order unity in most models. Two benchmark models are used: in the first, referred to as {\em Model A}, the branching 
ratios of the new heavy vector boson to known fermions and gauge bosons are comparable, as in some extensions of the SM gauge group~\cite{Barger:1980ix}. 
In {\em Model B}, fermionic couplings to the new heavy vector boson are suppressed, as for example in a composite Higgs model~\cite{Contino:2011np}. The regions of 
HVT parameter space studied correspond to the production of resonances with an intrinsic width that is narrow relative 
to the experimental resolution.  The latter is roughly 8\% of the resonance mass.  The sensitivity of the analysis to wider 
resonances is not tested.

Searches for $VH$ resonances, $V'$, have recently been performed by the ATLAS and CMS collaborations.  The ATLAS searches (using leptonic $V$ decays)
based on data collected at $\sqrt{s}$ = 8 \tev\ set a lower limit at the 95\% confidence level (CL) on the $W'$ ($Z'$) mass at 
1.47 (1.36) \tev\ in HVT benchmark {\em Model A} with $g_V=1$~\cite{Aad:2015yza}. 
Using the same decay modes, a search based on 3.2 \ifb\ of data collected at $\sqrt{s} = 13$ $\tev$ set a 95\% CL lower limit on the $W'$ ($Z'$) mass at 1.75 (1.49)~\tev~\cite{ATLASVH2015}
in the HVT benchmark {\em Model A}.
For {\em Model B} the corresponding limits are 2.22 (1.58)~\tev.
Searches by the CMS Collaboration at $\sqrt{s}$ = 8 \tev\ in hadronic channels, based on HVT benchmark {\em Model B} with $g_V$ = 3, exclude heavy resonance masses 
below 1.6~\tev\ ($W' \to WH$), below 1.1~\tev\ and between 1.3~\tev\ and 1.5~\tev\ ($Z' \to ZH$),  and below 1.7~\tev\ (combined $V' \to VH$)~\cite{Khachatryan:2015bma} at the 95\% CL. Using the 
$W' \to WH \to \ell\nu b\bar{b}$ channel, CMS excludes new heavy vector bosons with masses up to 1.5~\tev\ in the same context~\cite{Khachatryan:2016yji}.
The CMS Collaboration also carried out a search for a narrow resonance decaying to $ZH$ in the $q\bar{q}\tau^+\tau^-$ final state, 
setting limits on the $Z'$ production cross-section~\cite{Khachatryan:2015ywa}.  Searches for heavy resonances in HVT models have also been carried out in the hadronic $WW$/$WZ$/$ZZ$ channels by the ATLAS experiment at $\sqrt{s}=13$ \tev\ with 3.2~\ifb\ of data~\cite{ATLASVV2015}.  For {\em Model B}, a new gauge boson with mass below 2.6~\tev\ is excluded at the 95\%~CL.
The CMS Collaboration combined ~\cite{Sirunyan:2017nrt}
diboson resonance searches at $\sqrt{s}$ = 8 and 13~\tev~\cite{Sirunyan:2017nrt}, setting lower limits for $W'$ and $Z'$ 
singlets at 
2.3 \tev and for a triplet at 2.4 \tev.  As this Letter was being finalized, the CMS Collaboration 
released~\cite{Sirunyan:2017wto} a
search in the same final state as studied in this Letter, using 36~\ifb\ of data collected at $\sqrt{s}$ = 13~\tev.  For 
{\em Model B}, a $W'$ boson with mass below 2.45~\tev\ and between 2.78~\tev\ and 3.15~\tev\ is excluded at the 95\%~CL.  
For a $Z'$ boson, masses below 1.19~\tev\ and between 1.21~\tev\ and 2.26~\tev\ are excluded at the 95\%~CL.

\section{ATLAS detector}
\label{sec:detector}


The ATLAS detector~\cite{Aad:2008zzm} is a general-purpose particle detector used to investigate 
a broad range of physics processes. It includes inner tracking devices surrounded by a 2.3 m diameter superconducting 
solenoid, electromagnetic and hadronic calorimeters and a muon spectrometer with a toroidal magnetic field. 
The inner detector consists of a high-granularity silicon pixel detector, including the insertable 
B-layer~\cite{IBL} installed after Run 1 of the LHC, a silicon strip detector, and a straw-tube tracker. 
It is immersed in a 2 T axial magnetic field and provides precision tracking of charged particles with 
pseudorapidity $|\eta| <$ 2.5.\footnote{ATLAS uses a right-handed coordinate system with its origin 
at the nominal interaction 
point (IP) in the centre of the detector and the $z$-axis along the beam pipe. The $x$-axis points 
from the IP to the centre of the LHC ring, and the $y$-axis points upward. Cylindrical coordinates 
$(r,\phi)$ are used in the transverse plane, $\phi$ being the azimuthal angle around the $z$-axis. 
The pseudorapidity is defined in terms of the polar angle $\theta$ as $\eta=-\ln\tan(\theta/2)$. 
The rapidity is defined relative to the beam axis as $y = 1/2\ln((E + p_z)/(E - p_z)) $. } 
The calorimeter system consists 
of finely segmented sampling calorimeters using lead/liquid-argon for the 
detection of electromagnetic (EM)
showers up to $|\eta| <$ 3.2, and copper or tungsten/liquid-argon for 
electromagnetic and hadronic showers for 1.5 $< |\eta| <$ 4.9.  In the central 
region ($|\eta| <$ 1.7), a steel/scintillator hadronic calorimeter is used.
Outside the calorimeters, the muon system incorporates multiple layers of
trigger and tracking chambers within a magnetic field produced by a system of superconducting toroids,
enabling an independent precise measurement of muon track momenta for $|\eta| <$ 2.7.
A dedicated trigger system is used to select events~\cite{ATL-DAQ-PUB-2016-001}. The first-level trigger is implemented in hardware and uses the calorimeter and muon detectors to reduce the accepted rate to 100 kHz. This is followed by a software-based high-level trigger, which reduces the accepted event rate to 1 kHz on average.

\section{Data and simulation samples}
\label{sec:samples}

This analysis uses \lumi\ fb$^{-1}$ of LHC $pp$ collisions at $\sqrt{s}=13$ \tev\ collected in 2015 and 2016.  The data were collected during stable beam conditions with all relevant detector systems functional.  
Events were selected using a trigger that requires a single anti-$k_t$ jet~\cite{Cacciari:2008gp} with radius parameter $R=1.0$ (\largeR\ jet) with a transverse energy (\et) threshold of 360 (420) \gev\ in 2015 (2016).  The trigger requirement is $>99\%$ efficient for events passing the offline selection of a  \largeR\ jet with transverse momentum (\pt) $> 450$~\gev.

Signal and backgrounds from $t\bar{t}$ and $W/Z\ +$ jets production are modelled with Monte Carlo (MC) simulation.  While multijet MC events are used as a cross-check, the primary multijet background estimation is performed using data as described in Section~\ref{sec:backgrounds}. The signal is modelled using benchmark {\em Model A} with $g_V = 1$.  Results derived from this model can be directly applied to benchmark {\em Model B} by rescaling the relevant branching ratios.  The signal was generated with {\textsc Madgraph5}\_a{\textsc MC@NLO} 2.2.2~\cite{Alwall:2014hca} interfaced to \PYTHIA8.186~\cite{Sjostrand:2007gs} for parton shower and hadronization, with the NNPDF2.3 next-to-leading order (NLO) parton distribution function (PDF) set~\cite{Ball:2012cx} and a set of tuned parameters called the ATLAS A14 tune~\cite{ATL-PHYS-PUB-2014-021} for the underlying event.  The Higgs boson mass was set to 125.5~\gev, and Higgs decays to both $b\bar{b}$ and $c\bar{c}$, assuming the SM branching ratios, were included in the simulation.  The $V^{\prime} \to VH \to q \bar{q}^{(\prime)} (b\bar{b}+c\bar{c})$ signal cross-section in {\em Model B}  ranges from 110~fb (203~fb) for neutral (charged) resonances with a mass of 1~\tev, down to  0.09~fb (0.19~fb) for neutral (charged) resonances with a mass of 3.8~\tev.  Samples were generated in steps of 100 \gev\ or 200 \gev\ up to 4 \tev.

The $t\bar{t}$\, background samples were generated with \textsc{Powheg-Box} v2~\cite{Frixione:2002ik} with the CT10 PDF set~\cite{Lai:2010vv}, interfaced with \PYTHIA 6.428~\cite{Sjostrand:2006za} and the Perugia 2012 tune for the parton shower~\cite{Skands:2010ak} using the CTEQ6L1 PDF set~\cite{Pumplin:2002vw}.   The cross-section of the $t\bar{t}$ process is normalized to the result of a quantum chromodynamics (QCD) calculation at next-next-to-leading order and log (NNLO+NNLL), as implemented in \textsc{Top++} 2.0~\cite{Czakon:2011xx}. The \POWHEG \textsc{hdamp} parameter~\cite{ATL-PHYS-PUB-2014-005} was set to the top quark mass, taken to be $m_{t} = 172.5$~\gev. The $W$+jets and $Z$+jets background samples were generated with \SHERPA 2.1.1~\cite{Gleisberg:2008ta}  interfaced with the CT10 PDF set. Matrix elements of up to four extra partons were calculated at leading order in QCD. Only the hadronic decays of the $W$ and $Z$ bosons were included.  For studies with simulated multijet events, the MC samples were generated with \PYTHIA 8.186~\cite{Sjostrand:2007gs}, with the NNPDF2.3 NLO PDF and the ATLAS A14 tune.  The background from SM diboson and $VH$ production is negligible and therefore not considered.

For all simulated events, except those produced using \SHERPA, \textsc{EvtGen} v1.2.0~\cite{Lange:2001uf} was used to model the properties of bottom and charm hadron decays. 
The detector response was simulated with \textsc{Geant 4}~\cite{Agostinelli:2002hh, Aad:2010ah} and the events were processed with the same reconstruction software as that used for data.  
All simulated samples include the effects
due to multiple $pp$ interactions per bunch-crossing (pile-up).

\section{Event reconstruction}
\label{sec:eventReco}
Collision vertices are reconstructed requiring a minimum of two tracks each with transverse momentum $\pt > 0.4$ \gev. The primary vertex is chosen to be the vertex with the largest $\sum \pt^2$, where the sum extends over all tracks associated with the vertex.

The identification and reconstruction of hadronically decaying gauge boson and Higgs boson candidates is performed with the anti-$k_t$ jet clustering algorithm with $R$ parameter equal to 1.0.  These \largeR\ jets~\cite{Aad:2013gja} are reconstructed from locally calibrated topological clusters~\cite{PERF-2014-07} of calorimeter energy deposits. To mitigate the effects of pile-up and soft radiation, the \largeR\ jets are trimmed~\cite{Krohn2010}:  the jet constituents are reclustered into subjets using the $k_t$ algorithm~\cite{Catani:1993hr} with $R=0.2$, removing those with $\pt^{\textrm{subjet}} / \pt^{\textrm{jet}} < 0.05$, where $\pt^{\textrm{subjet}}$ is the transverse momentum of the subjet and $\pt^{\textrm{jet}}$ is the transverse momentum of the original \largeR\ jet. In order to improve on the limited angular resolution of the calorimeter, the combined mass of a \largeR\ jet is computed using a combination of calorimeter and tracking information~\cite{ATLAS-CONF-2016-035}. The combined mass is defined as:

\[
m_{J} \equiv w_{\textrm calo} \times \mJcalo + w_{\textrm track} \times \Big( \mJtrack \frac{\pTcalo}{ \pTtrack} \Big),
\]
where $\mJcalo$ ($\pTcalo$) is the calorimeter-only estimate of the jet mass (\pt), and $\mJtrack$ ($\pTtrack$) is the jet mass (\pt) estimated via tracks with $\pt > 0.4$ \gev\  associated with the \largeR\ jet using ghost association\footnote{In this method, the \largeR\ jet algorithm is rerun with both the four-momenta of tracks, modified to have infinitesimally small momentum (the ``ghosts''), and all topological energy clusters in the event as potential constituents of jets. As a result, the presence of tracks does not alter the \largeR\ jets already found and their association with specific \largeR\ jets is determined by the jet algorithm.}~\cite{Cacciari:2007fd}.  To correct for the missing neutral component in the track-based measurement,  $\mJtrack$ is scaled by the ratio of calorimeter to track \pt\ estimates.  The weighting factors $w_{\textrm calo}$ and $w_{\textrm track}$ are $\pTcalo$-dependent functions of the calorimeter- and track-based jet mass resolutions used to optimize the combined mass resolution.

Track jets clustered using the anti-$k_t$ algorithm with $R=0.2$ are used to aid the identification of $b$-hadron candidates from the Higgs boson decay~\cite{ATL-PHYS-PUB-2015-035}.  Track jets are built from charged particle tracks with $\pt > 0.4$ \gev\ and $|\eta| < 2.5$ that satisfy a set of hit and impact parameter criteria to minimize the impact of tracks from pile-up interactions, and are required to have track jet $\pt > 10$ \gev, $|\eta| < 2.5$, and at least two tracks clustered in the track jet. Track jets are matched with \largeR\ jets using ghost association.  The identification of $b$-hadrons relies on a multivariate tagging algorithm~\cite{Calandri:2160731} which combines information from several vertexing and impact parameter tagging algorithms applied to a set of tracks in a region of interest around each track jet axis. The $b$-tagging requirements result in an efficiency of 77\% for track jets containing $b$-hadrons, and a misidentification rate of  $\sim 2\%$ ($\sim 24\%$) for light-flavour (charm) jets, as determined in a sample of simulated $t\bar{t}$ events.  For MC samples the tagging efficiencies are corrected to match those measured in data~\cite{Aad:2015ydr}.

Muons are reconstructed by combining tracks in the inner detector and the muon system, and are required to satisfy ``Tight'' muon identification criteria~\cite{Aad:2016jkr}. The four-momentum of the closest muon candidate with $\pt\ > 4$~\gev\ and $|\eta| < 2.5$ that is within $\Delta R = \sqrt{(\Delta \eta)^2 + (\Delta \phi)^2} = 0.2$ of a track jet is added to the calorimeter jet four-momentum to partially account for the energy carried by muons from semileptonic $b$-hadron decays.  This muon correction results in a \mbox{$\sim5\%$} resolution improvement for Higgs boson candidate jets (defined in Section~\ref{sec:eventSelection})~\cite{XbbCONF2016}.  Electrons are reconstructed from inner detector and calorimeter information, and are required to satisfy the ``Loose'' likelihood selection~\cite{Aad:2014fxa}.

Leptons (electrons and muons, $\ell$) are also used in a ``veto'' to ensure the orthogonality of the analysis selection with respect to other heavy $VH$ resonance searches in non-fully hadronic final states.  The considered leptons have $\pt\ > 7$ \gev, $|\eta| < 2.5\ (2.47)$ for muons (electrons), and their associated tracks must have $|d_0| / \sigma_{d_0} < 3\ (5)$ and $|z_0 \sin \theta| < 0.5$ mm, where $d_0$ is the transverse impact parameter with respect to the beam line, $\sigma_{d_0}$ is the uncertainty on $d_0$, and $z_0$ is the distance between the longitudinal position of the track along the beam line at the point where $d_0$ is measured and the longitudinal position of the primary vertex.  Leptons are also required to satisfy an isolation criterion, whereby the ratio of the \pt\ sum of all tracks  with $\pt > 1$ \GeV\ (excluding the lepton's) within a cone around the lepton (with radius dependent on the lepton \pt) to the lepton momentum must be less than a \pt- and $|\eta|$-dependent threshold $I_0$.  The value of $I_0$ is chosen such that a constant efficiency of 99\% as a function of \pt and $|\eta|$ is obtained for leptons in events with identified $Z \to \ell\ell$ candidates.

The missing transverse momentum ($\vec{E}^{\textrm miss}_{\textrm T}$) is calculated as the negative
vectorial sum of the transverse momenta of all the muons, electrons, calorimeter jets with $R=0.4$, and any inner-detector
tracks from the primary vertex not matched to any of these objects~\cite{EtmissPUB}. The magnitude of the $\vec{E}^{\textrm miss}_{\textrm T}$ is denoted by \met.

\section{Event selection}
\label{sec:eventSelection}

Events selected for this analysis must contain at least two large-$R$ jets with $|\eta| < 2.0$ and invariant mass $m_{J} > 50$~\gev, and cannot have any lepton candidate passing the veto for leptons. The leading and subleading \pt\ large-$R$ jets must have $\pt$ greater than $450$~\gev\ and 250~\gev, respectively.
The two leading \pt\ large-$R$ jets are assigned to be the Higgs and vector boson candidates, and the invariant mass of the individual jets is used to determine the boson type; the \largeR\ jet with the larger invariant mass is assigned to be the Higgs boson candidate jet ($H$-jet), while the smaller invariant mass \largeR\ jet is assigned as the vector boson candidate jet ($V$-jet). In signal MC simulation, this procedure results in 99\% correct assignment after the full signal region selections described below. Furthermore, the absolute value  of the rapidity difference, $|\Delta y_{12}|$, between the two leading \pt\ \largeR\ jets must be less than $1.6$, exploiting the more central production of the signal compared to the multijet background.  To ensure orthogonality with the $ZH$ resonance search in which the $Z$ boson decays to neutrinos, events are rejected if they have $\met > 150$ \gev\ and $\Delta \phi (\vec{E}^{\textrm miss}_{\textrm T},\ H\textrm{-jet}) > 120$ degrees.

The $H$-jet is further required to satisfy mass and $b$-tagging criteria consistent with expectations from a Higgs boson decaying to $b\bar{b}$~\cite{ATL-PHYS-PUB-2015-035}.  The $H$-jet mass, $m_{J, H}$, must satisfy  $75$ $\gev < m_{J, H} < 145$ \gev, which is $\sim90\%$ efficient for Higgs boson jets.  The number of ghost associated $b$-tagged track jets is then used to categorize events. The $H$-jets with either one or two $b$-tagged track jets, amongst the two leading \pt\ associated track jets, are used in this analysis. The $H$-jets with one associated $b$-tagged track jet are not required to have two associated track jets.  The Higgs tagging efficiency, defined with respect to jets that are within $\Delta R = 1.0$ of a truth Higgs boson and its decay $b$-hadrons,  for doubly- (singly-) $b$-tagged $H$-jets is $\sim 40\%$ ($\sim 75\%$) for $H$-jets with $\pt \approx 500$~\gev\ and  $\sim 25\%$ ($\sim 65\%$) for $H$-jets with $\pt \approx 900$~\gev~\cite{XbbCONF2016}.  The rejection factor for jets from multijet production is $\sim 600$ ($\sim 50$) for double (single) tags.

The $V$-jet must satisfy mass and substructure criteria consistent with a $W$- or $Z$-jet using a 50\% efficiency working point, similar to the ``Medium'' working point in Ref.~\cite{BosonTagPreRec}.  To be considered a $W$ ($Z$) candidate, the $V$-jet must have a mass $m_{J, V}$ within a \pt-dependent mass window which varies between $m_{J, V} \in [67,\ 95]\ ([75,\ 107])$~\gev\ for jets with $\pt \approx 250$ \gev, and $m_{J, V} \in [60,\ 100]\ ([70,\ 110])$~\gev\ for jets with $\pt \approx 2500$ \gev.  The jet must also satisfy a \pt-dependent $D_2$~\cite{Larkoski:2014gra,Larkoski:2015kga} selection (with $\beta=1$) which depends on whether the candidate is a $W$ or a $Z$ boson, as described in Ref.~\cite{BosonTagPreRec}.  The variable $D_2$ exploits two- and three-point energy correlation functions to tag boosted objects with two-prong decay structures.  The $V$-jet tagging efficiency is $~\sim50\%$ and constant in $V$-jet \pt, with a misidentification rate for jets from multijet production of $\sim 2\%$.

Four signal regions (SRs) are used in this analysis.  They differ by the number of $b$-tagged track jets associated to the $H$-jet and by whether the $V$-jet passes a $Z$-tag or $W$-tag selection.  The ``1-tag'' and ``2-tag'' SRs require exactly one and two $b$-tagged track jets associated to the $H$-jet, respectively.  The 2-tag signal regions provide better sensitivity for resonances with masses below $\sim 2.5$ TeV.  Above $2.5$ TeV the 1-tag regions provide higher sensitivity because the Lorentz boost of the Higgs boson is large enough to merge the fragmentation products of both $b$-quarks into a single track jet.  Events in which the $V$-jet passes a $Z$-tag constitute the $ZH$ signal regions, while events in which the $V$-jet passes a $W$-tag constitute the $WH$ signal regions. While the 1-tag and 2-tag signal regions are orthogonal regardless of the $V$-jet tag, the $WH$ and $ZH$ selections are not orthogonal within a given $b$-tag category. The overlap between the $WH$ and $ZH$ selections in the signal regions is approximately 60\%.

The final event requirement is that the mass of the candidate resonance built from the sum of the $V$-jet and $H$-jet candidate four-momenta, $m_{JJ}$, must be larger than 1 \tev.  This requirement ensures full efficiency for the trigger and jet \pt\ requirements for events passing the full selection.  The full event selection can be found in Table~\ref{tab:selection}.  The expected selection efficiency for both $WH$ and $ZH$ resonances decaying to $q \bar{q}^{(\prime)} (b\bar{b}+c\bar{c})$ with a mass of 2 (3) \tev\ in the  HVT benchmark {\em Model B} is $\sim 30\%$ ($\sim 20\%$).

\begin{table}[htbp!]
\scriptsize
\centering
\caption{Summary of event selection criteria. The selection efficiency for HVT benchmark {\em Model B} is shown
for $WH$ resonances.  It is very similar for $ZH$ resonances.}
\begin{tabular}{l|c|c}
\hline
Selection & Description & $m$ = 2 \tev\ $WH$ signal efficiency [\%] \\
\hline

Large-$R$ jet selection &  ${\pt}^{\textrm lead}>450$ \gev, ${\pt}^{\textrm sublead}>250$ \gev, $|\eta|<2.0$, $m_{J} > 50$~\gev & 83.8 \\
Lepton veto & Remove events with leptons & 83.0\\
Rapidity difference & $|\Delta y_{12}|<1.6$ & 73.3 \\
\met\ veto & Remove events with $\met > 150$ \gev\ and $\Delta \phi (\vec{E}^{\textrm miss}_{\textrm T},\ H\textrm{-jet}) > 120$ degrees & 68.3\\
$V/H$ assignment & Larger mass jet is $H$-jet, smaller mass jet is $V$-jet & 68.3 \\
$W/Z$-tagging & Mass window + $D_{2}$ selection & 36.3 \\
Dijet mass & $m_{JJ} > 1$ \tev & 36.3 \\
\hline
Signal region & $WH$ 1-tag &  15.0 \\
Signal region  & $WH$ 2-tag &  12.5 \\
\hline
\end{tabular}
\label{tab:selection}
\end{table}

\section{Background estimation}
\label{sec:backgrounds}

After the selection of 1-tag and 2-tag events, $\sim 90\%$ of the background in the signal regions originates from multijet events.    The remaining $\sim 10\%$ is predominantly \ttbar\ with a small contribution from $V$+jets ($ \lesssim 1\%$).  The multijet background is modelled directly from data, while other backgrounds are estimated from MC simulation.

Multijet modelling starts from the same trigger and event selection as described above, but the $H$-jet is required to have zero associated $b$-tagged track jets.  This 0-tag sample, which consists of multijet events at the $\sim  99\%$  level, is used to model the kinematics of the multijet background in the 1-tag and 2-tag SRs.  To keep the 0-tag region kinematics close to the 1- and 2-tag regions, $H$-jets in 0-tag events must contain at least one (two) associated track jets when modelling the  1(2)-tag signal region.  

The 0-tag sample is normalized to the 1-tag and 2-tag samples and corrected for kinematic differences  with respect to the signal regions.  These kinematic differences arise from the $b$-tagging efficiency variations as a function of \pt\ and $|\eta|$ and because different multijet processes,  in terms of quark, gluon, and heavy-flavour content, contribute different fractions to the 0-, 1-, and 2-tag samples.    

The 0-tag sample is normalized to the 1- and 2-tag samples, separately, using a signal-free high mass sideband of the $H$-jet defined by $145\, \gev < m_{J, H} < 200$ \gev.  This sideband (SB), illustrated in Figure~\ref{fig:cartoons}, is orthogonal to the signal region and has similar expected event yield to the signal region.  
 The normalization of the multijet events is set by scaling the number of events in each region of the 0-tag sample by

\begin{equation}
\mu_{\textrm{Multijet}}^{1(2)\textrm{-tag}} = \frac{N_{\textrm{Multijet}}^{1(2)\textrm{-tag}}}{N_{\textrm{Multijet}}^{0\textrm{-tag}}} =  \frac{ N_{\textrm{data}}^{1(2)\textrm{-tag}} - N_{\ttbar}^{1(2)\textrm{-tag}} - N_{V+\textrm{jets}}^{1(2)\textrm{-tag}} }{ N_{\textrm{data}}^{0\textrm{-tag}} - N_{\ttbar}^{0\textrm{-tag}} - N_{V+\textrm{jets}}^{0\textrm{-tag}} },
\end{equation} 
where $N_{\textrm{data}}^{0/1/2\textrm{-tag}}$, $N_{\ttbar}^{0/1/2\textrm{-tag}}$ and $N_{V+\textrm{jets}}^{0/1/2\textrm{-tag}}$ are the numbers of events observed in data, and predicted from \ttbar and $V$+jets MC simulation in the 0-, 1-, or 2-tag SB samples, respectively.  As the selection of track jets for $H$-jets in 0-tag events differs when modelling the 1-tag and 2-tag regions (as stated above),  $N_{\textrm{Multijet}}^{0\textrm{-tag}}$ differs between estimates of the $\mu_{\textrm{Multijet}}^{1\textrm{-tag}}$ and $\mu_{\textrm{Multijet}}^{2\textrm{-tag}}$. 

\begin{figure}[!htbp]
\begin{center}
\includegraphics[width=0.40\textwidth]{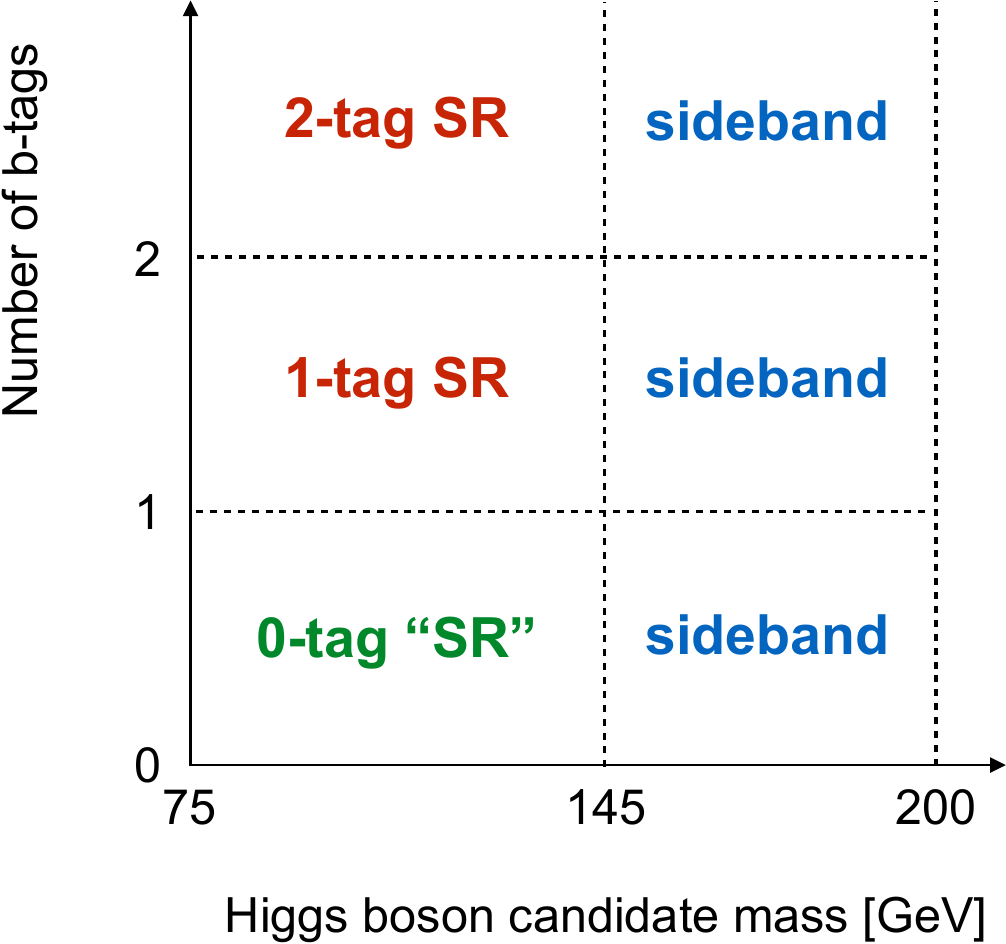}
\includegraphics[width=0.40\textwidth]{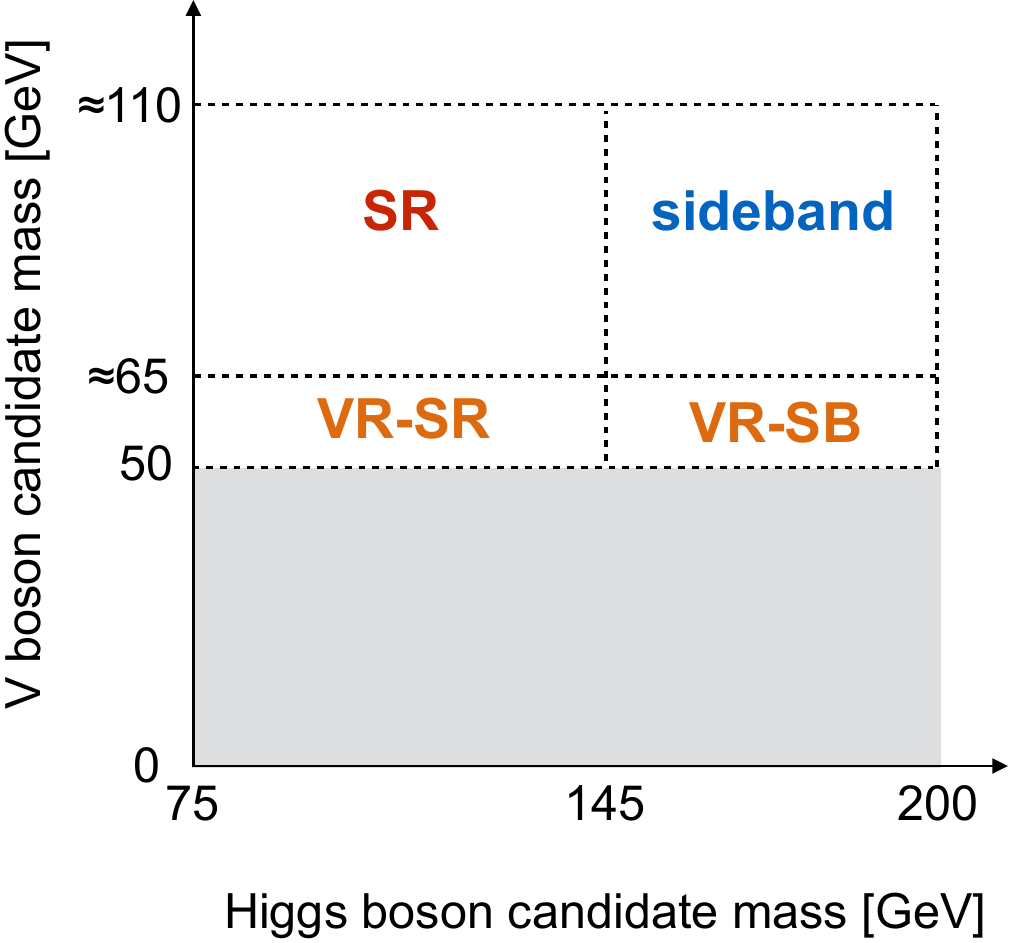}
\caption{Illustration of the sideband and validation regions, showing orthogonal slices through the space defined by the masses of the two boson candidates and the number of $b$-tags.}
\label{fig:cartoons}
\end{center}
\end{figure}

Kinematic corrections to the multijet background template are applied by reweighting events from the 0-tag sample.   This is performed only for the 2-tag sample, as the modelling of the multijet background in the 1-tag SB and validation regions (described below and depicted in Figure~\ref{fig:cartoons}) without reweighting is observed to be adequate.  The weights are derived in the SB region, from third-order polynomial fits to the ratio of the total background model to data in two distributions that are sensitive to kinematic and $b$-tagging efficiency differences between the 0-tag and 2-tag samples.  The variables are
the track jet \pt\ ratio, defined as $\pt^{\textrm lead}/(\pt^{\textrm lead}+\pt^{\textrm sublead})$, and $\pt^{\textrm sublead}$, both using the \pt\ distributions of the leading two \pt track jets associated to the $H$-jet.
The reweighting is performed using one dimensional distributions but is iterated so that correlations between the two variables are taken into account. After each reweighting iteration, the value of $\mu_{\textrm{Multijet}}^{1(2)\textrm{-tag}}$ is recomputed to ensure that the normalization is kept fixed.  No explicit uncertainties are associated with this reweighting as these are determined from comparison with validation regions, as described below.

Due to the small number of events in the background template in the high $m_{JJ}$ tail, the backgrounds are modelled by fitting between 1.2 and 4 \tev\ with power-law and exponential functions.  The multijet background in $m_{JJ}$ is modelled using the functional form

\begin{equation}
f_{\textrm{Multijet}}(x) = p_a (1-x)^{p_b} (1+x)^{p_c x},
\end{equation}
while the merged \ttbar\ and $V$+jets backgrounds are modelled using the functional forms
\begin{eqnarray}
f^{\textrm{1-tag}}_{\textrm Other}(x) = p_d (1-x)^{p_e} x^{p_f}, {\textrm and} \\
f^{\textrm{2-tag}}_{\textrm Other}(x) = p_g \mathrm{e}^{- p_h x}
\end{eqnarray}
for the 1-tag and 2-tag samples respectively.  In these functional forms, $x = m_{JJ} / \sqrt{s}$, and $p_a$ through $p_h$ are parameters determined by the fit.  These functional forms are used as they can model changes in the power-law behaviour of the respective backgrounds between high and low masses. The exponential function is used for the 2-tag \ttbar\ and $V$+jets samples because it was found to model the tail of the distribution well and because a fit to the small statistics of the sample could not constrain a function with more parameters.  Fits are performed separately for the 1-tag and 2-tag background estimates, and separately for each background.

The background model is validated in the two regions denoted by VR-SR and VR-SB in Figure~\ref{fig:cartoons}, each also with two subregions.  In all of these, the $V$-jet is required to have mass $50$ $\gev < m_{J, V} < 70$ \gev\ but the $D_2$ selection is only applied in one of the subregions.
For the signal-region-like validation regions (VR-SR) the $H$-jet selection is unchanged, and for the sideband-like validation regions (VR-SB) the $H$-jet is required to have mass $145$ $\gev < m_{J, H} < 200$ \gev.  Both validation regions are kinematically similar to the signal regions but orthogonal to them (and each other).  

Table ~\ref{tab:pred} compares the observed data yields in the validation regions with the corresponding background estimates. The differences are used as estimators of the background normalization uncertainties, as described in Section~\ref{sec:systematics}.
The modelling of the $m_{JJ}$ distribution in the signal-region-like validation region is shown in Figure~\ref{fig:mVH_VRSR} for the 1-tag and 2-tag samples. The data are well described by the background model.  Other kinematic variables are generally well described.

\begin{table}[htbp!]
\begin{center}
\caption{The number of events in data and predicted background events in the sideband and validation regions. In the sideband, the data 
 and total background agree by construction.
  The uncertainties are statistical only.  
 Due to rounding the totals can differ from the sums of components.}
\begin{tabular}{lD{,}{\,\,}{-1}D{,}{\,\,}{-1}D{,}{\,\,}{-1}D{,}{\,\,}{-1}D{,}{\,\,}{-1}}
\toprule
2-tag  & \multicolumn{1}{c}{Sideband region} & \multicolumn{2}{c}{Validation region} & \multicolumn{2}{c}{Validation region} \\
sample & & \multicolumn{2}{c}{(Signal-region-like)} & \multicolumn{2}{c}{(Sideband-region-like)} \\
  & & \multicolumn{1}{c}{No $D_2$} & \multicolumn{1}{c}{With $D_2$} & \multicolumn{1}{c}{No $D_2$} & \multicolumn{1}{c}{With $D_2$} \\
\midrule
Multijet & 1410, \pm 10 & 13700, \pm 20 & 875, \pm 5  & 7150, \pm 10 & 455, \pm 5 \\
$t\bar{t}$ & 220, \pm 10 & 115, \pm 10  & 12, \pm 3 & 250, \pm 15 & 26, \pm 4  \\
$V$+jets &  35, \pm 15 & 250, \pm 30 & 14, \pm 6  & 30, \pm 10 & 3, \pm 3  \\
\midrule
Total & 1670, \pm 20 & 14050, \pm 35 & 900, \pm 8 & 7430, \pm 20 & 485,  \pm 6  \\
\midrule
Data & 1667, & 15013, & 934, & 7200, & 426, \\
\toprule
 & & & \\
1-tag  & \multicolumn{1}{c}{Sideband region} & \multicolumn{2}{c}{Validation region} & \multicolumn{2}{c}{Validation region} \\
sample &  & \multicolumn{2}{c}{(Signal-region-like)} & \multicolumn{2}{c}{(Sideband-region-like)} \\
  & & \multicolumn{1}{c}{No $D_2$} & \multicolumn{1}{c}{With $D_2$} & \multicolumn{1}{c}{No $D_2$} & \multicolumn{1}{c}{With $D_2$} \\
\midrule
Multijet & 12350, \pm 50 & 138500, \pm 160 & 8820, \pm 40  & 62600, \pm 100 & 3970, \pm 30  \\
$t\bar{t}$ & 2200, \pm 30  & 1030, \pm 30 & 115, \pm 7  & 1700, \pm 35 & 210, \pm 10 \\
$V$+jets & 300, \pm 40  & 1480, \pm 90  &  120, \pm 25 & 420, \pm 50 & 35, \pm 13 \\
\midrule
Total & 15000, \pm 75  & 140900, \pm 190 & 9050, \pm 50  & 64700, \pm 120 & 4200, \pm 30 \\
\midrule
Data & 14973,  & 135131, & 8685, & 66896, & 4418,\\
\bottomrule
\end{tabular}
\label{tab:pred} 
\end{center}
\end{table}

\begin{figure}[!htbp]
\begin{center}
\includegraphics[width=0.49\textwidth]{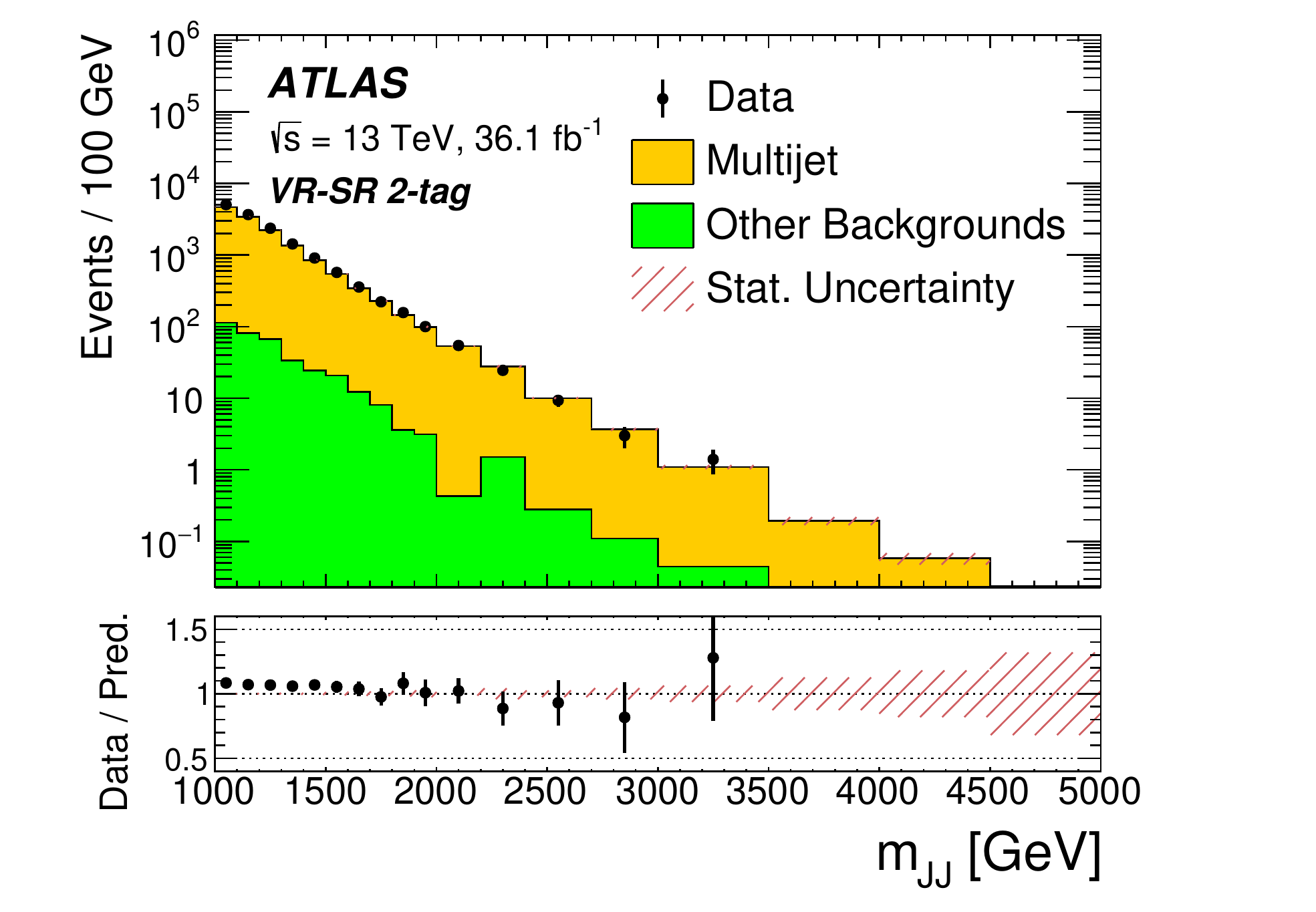}
\includegraphics[width=0.49\textwidth]{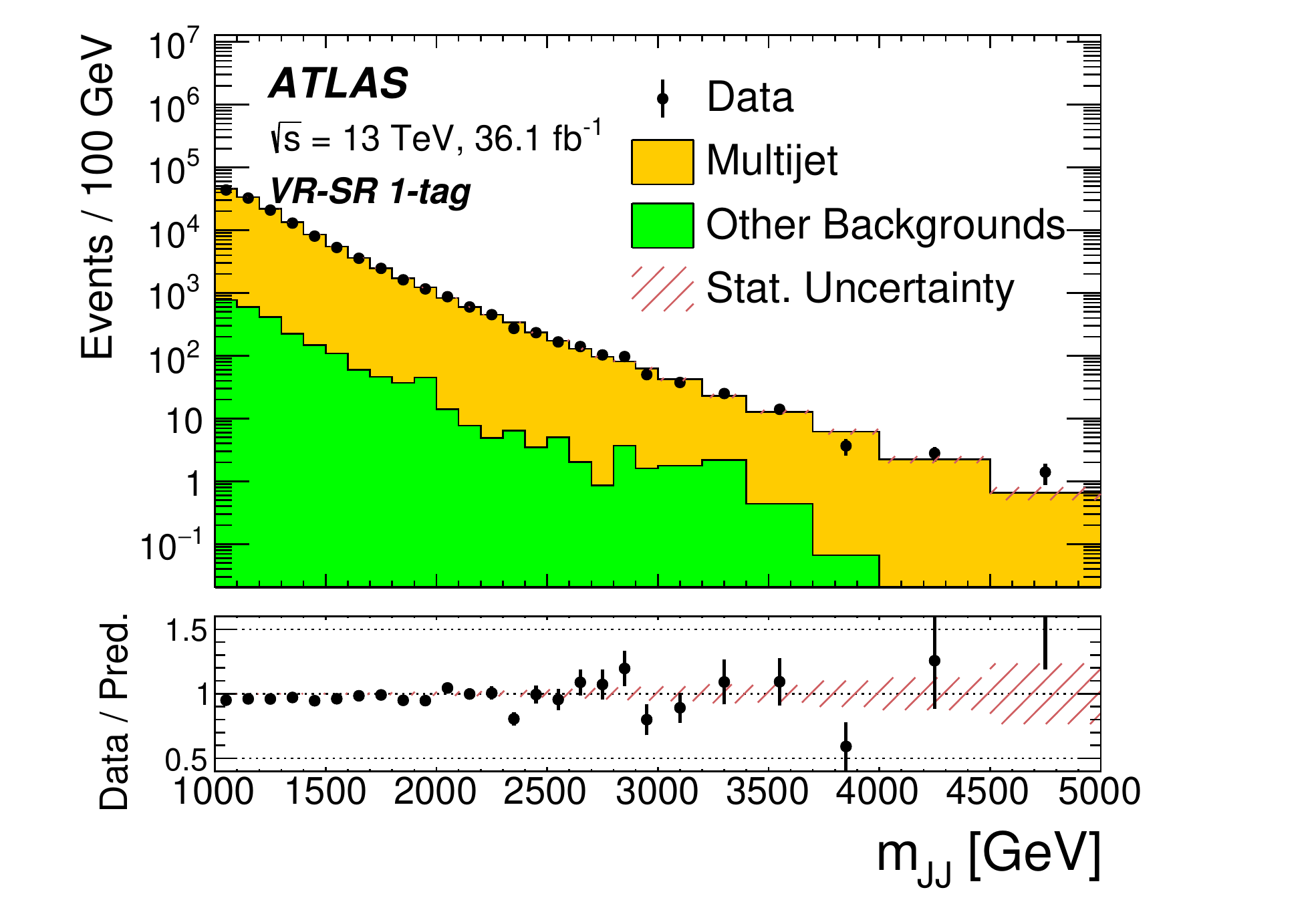}
\caption{The $m_{JJ}$ distribution in the signal-region-like validation region in the (left) 2-tag (right) 1-tag samples, compared to the predicted background.  The uncertainty band corresponds to the statistical uncertainty on the 
multijet model.}
\label{fig:mVH_VRSR}
\end{center}
\end{figure}

\section{Systematic uncertainties}
\label{sec:systematics}

The preliminary uncertainty on the combined 2015 and 2016 integrated luminosity is \lumiErr\%. It is derived, following a methodology similar to that detailed in Ref.~\cite{ATLASlumi8TeV}, from a preliminary calibration of the luminosity scale using $x$--$y$ beam-separation scans performed in 2015 and 2016.

Experimental systematic uncertainties affect the signal as well as the \ttbar and $V$+jets backgrounds estimated from MC simulation. The systematic uncertainties related to the scales of the \largeR\ jet \pt, mass and $D_{2}$ are of the order of 2\%, 5\% and 3\%, respectively. They are derived 
following the technique described in Ref.~\cite{Aad:2013gja}. 
The impacts of the uncertainties on the resolutions of each of these \largeR\ jet observables are evaluated by smearing the jet observable according to the systematic uncertainties of the resolution measurement~\cite{Aad:2013gja,BosonTagPreRec}.  A 2\% absolute uncertainty is assigned to the \largeR\ jet \pt, and to the mass and $D_2$ resolutions relative 20\% and 15\% uncertainties are assigned, respectively.  The uncertainty in the $b$-tagging efficiency for track jets is based on the uncertainty in the measured tagging efficiency for $b$-jets in data following the methodology used in Ref.~\cite{Aad:2015ydr}. This is measured as a function of $b$-jet \pt and ranges between 2\% and 8\% for track jets with $\pt < 250$ \gev.  For track jets with \pt~$> 250$ \gev\ the uncertainty in the tagging efficiencies is extrapolated using MC simulation~\cite{Aad:2015ydr} and is approximately 9\% for track jets with \pt\ $> 400$ \gev.  A 30\% normalization uncertainty is applied to the \ttbar background based on the ATLAS \ttbar differential cross-section measurement~\cite{Aad:2015hna}.  Due to the small contribution of the $V$+jets background, no corresponding uncertainty is considered.

Systematic uncertainties in the normalization and shape of the data-based multijet background model are assessed from the validation regions. The background normalization predictions in the validation regions agree with the observed data to within $\pm 5\%$ in the 1-tag sample and $\pm 13\%$ in the 2-tag sample.  These differences are taken as the uncertainties in the predicted multijet yield. The shape uncertainty is derived by taking the ratio of the predicted background to the observed data after fitting both to a power law.  This is done separately for the 1-tag and 2-tag samples. The larger of the observed shape differences in the VR-SR and VR-SB is taken as the shape uncertainty. Separate shape uncertainties are estimated for $m_{JJ}$ above and below 2~\tev\ in order to allow for independent shape variations in the bulk and tail of the $m_{JJ}$ distribution in the final statistical analysis.

An additional uncertainty in the shape of the multijet background prediction
is assigned by fitting a variety of empirical functions
designed to model power-law behaviour to the 0-tag $m_{JJ}$ distribution, as described in Ref.~\cite{Aad:2015mzg}. 
The largest difference between the nominal and alternative fit functions is taken as a systematic uncertainty.  Similarly, the fit range of the nominal power-law function is varied, and the largest difference between the nominal and alternative fit ranges is taken as a systematic uncertainty.  

The impact of the main systematic uncertainties on event yields is summarized in Table~\ref{tab:Syst}.

\begin{table}[ht!]
\begin{center} 
\small
\caption{Summary of the main post-fit systematic uncertainties (expressed as a percentage of the yield) 
 in the background and signal event yields in the 1-tag and 2-tag signal regions.
 The values for the jet energy scale and $b$-tagging efficiency uncertainties represent the sum in 
  quadrature of the values from the dominant components.
 The jet energy scale, jet mass resolution, $b$-tagging efficiency and luminosity do not apply to the 
 multijet contribution, which is determined from data.
 Uncertainties are provided for a resonance mass of 2~\TeV{} in the context of the
 HVT {\em Model B},  for both $V^\prime \to ZH$ and $V^\prime \to WH$ resonances.  }
\begin{tabular}{lcD{.}{.}{-1}cD{.}{.}{-1}}
\toprule
 & \multicolumn{2}{c}{$ZH$ 2-tag yield variation [\%]} & \multicolumn{2}{c}{$ZH$ 1-tag yield variation [\%]} \\
Source & Background & \multicolumn{1}{c}{HVT {\em Model B} $Z^\prime$(2 \tev)}  &  Background  & \multicolumn{1}{c}{HVT {\em Model B} $Z^\prime$(2 \tev)} \\
\midrule
Luminosity &  0.2  &  \lumiErr & 0.3 &  \lumiErr \\
Jet energy scale & 2.2 & 7.0 & 1.2 & 7.4 \\  
Jet mass resolution       & 0.6 & 9.5 & 0.4 & 8.5 \\  
$b$-tagging & 1.6 & 10 & 0.5 & 15 \\ 
$t\bar{t}$ normalization & 1.8  & - & 2.5  & - \\   
\midrule
Multijet normalization & 4.7 & - & 2.8 & - \\   
\toprule
& & & & \\

 & \multicolumn{2}{c}{$WH$ 2-tag yield variation [\%]} & \multicolumn{2}{c}{$WH$ 1-tag yield variation [\%]} \\
Source & Background & \multicolumn{1}{c}{HVT {\em Model B} $W^\prime$(2 \tev)} &  Background  & \multicolumn{1}{c}{HVT {\em Model B} $W^\prime$(2 \tev)} \\
\midrule
Luminosity &  0.2  &  \lumiErr & 0.3 &  \lumiErr \\
Jet energy scale & 2.4 & 5.7 & 0.8 & 5.6 \\  
Jet mass resolution       & 1.2 & 11 & 0.3 & 10 \\  
$b$-tagging & 1.6 & 10 & 0.4 & 15 \\ 
$t\bar{t}$ normalization & 1.9  & - & 2.5  & - \\   
\midrule
Multijet normalization & 4.3 & - & 2.8 & - \\   
\bottomrule
\end{tabular}
\label{tab:Syst}
\end{center}
\end{table}

\section{Results}
\label{sec:result}

The results are interpreted using the statistical procedure described in Ref.~\cite{Aad:2012tfa} and references therein.
A test statistic based on the profile likelihood ratio~\cite{Cowan:2010js} is used
to test hypothesized values of $\mu$, the global signal strength factor, separately for each model considered.
The statistical analysis described below is performed using the $m_{JJ}$ distribution of the data observed in the signal regions.
The systematic uncertainties are modelled with 
Gaussian or log-normal constraint terms (nuisance parameters) in the definition of the likelihood function.
The data distributions from the 1-tag and 2-tag signal regions are used in the fit simultaneously, treating systematic uncertainties on the luminosity, jet energy scale, jet energy resolution, jet mass resolution and $b$-tagging as fully correlated between the two signal regions. Both the multijet normalization and shape uncertainties are treated as independent between the two signal regions. In addition, the multijet shape uncertainties for $m_{JJ}$ above and below 2~\tev\ are treated as independent. 
When performing the fit, the nuisance parameters are allowed to vary within their constraints to maximize 
the likelihood.  As a result of the fit, the multijet shape uncertainties are significantly 
reduced.  With the jet mass resolution, jet energy scale and multijet normalization, they have the 
largest impact on the search sensitivity.
Fits in the $WH$ and $ZH$ signal regions are performed separately.  The pre- and post-fit $m_{JJ}$ distributions in the signal regions are shown in Figure~\ref{fig:m2J_SR}.

\begin{figure}[htbp!]
\begin{center}
\includegraphics[width=0.49\textwidth]{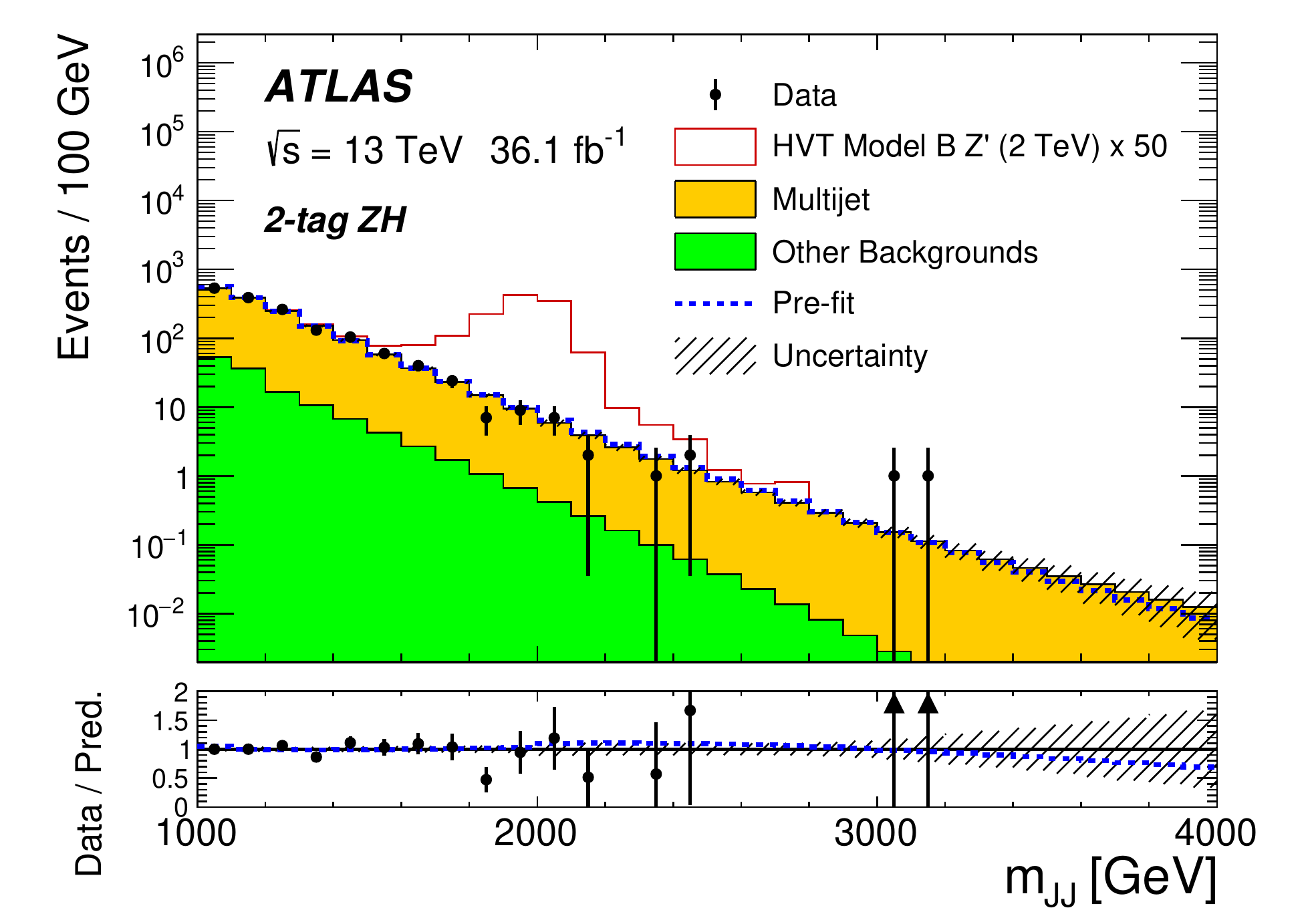} 
\includegraphics[width=0.49\textwidth]{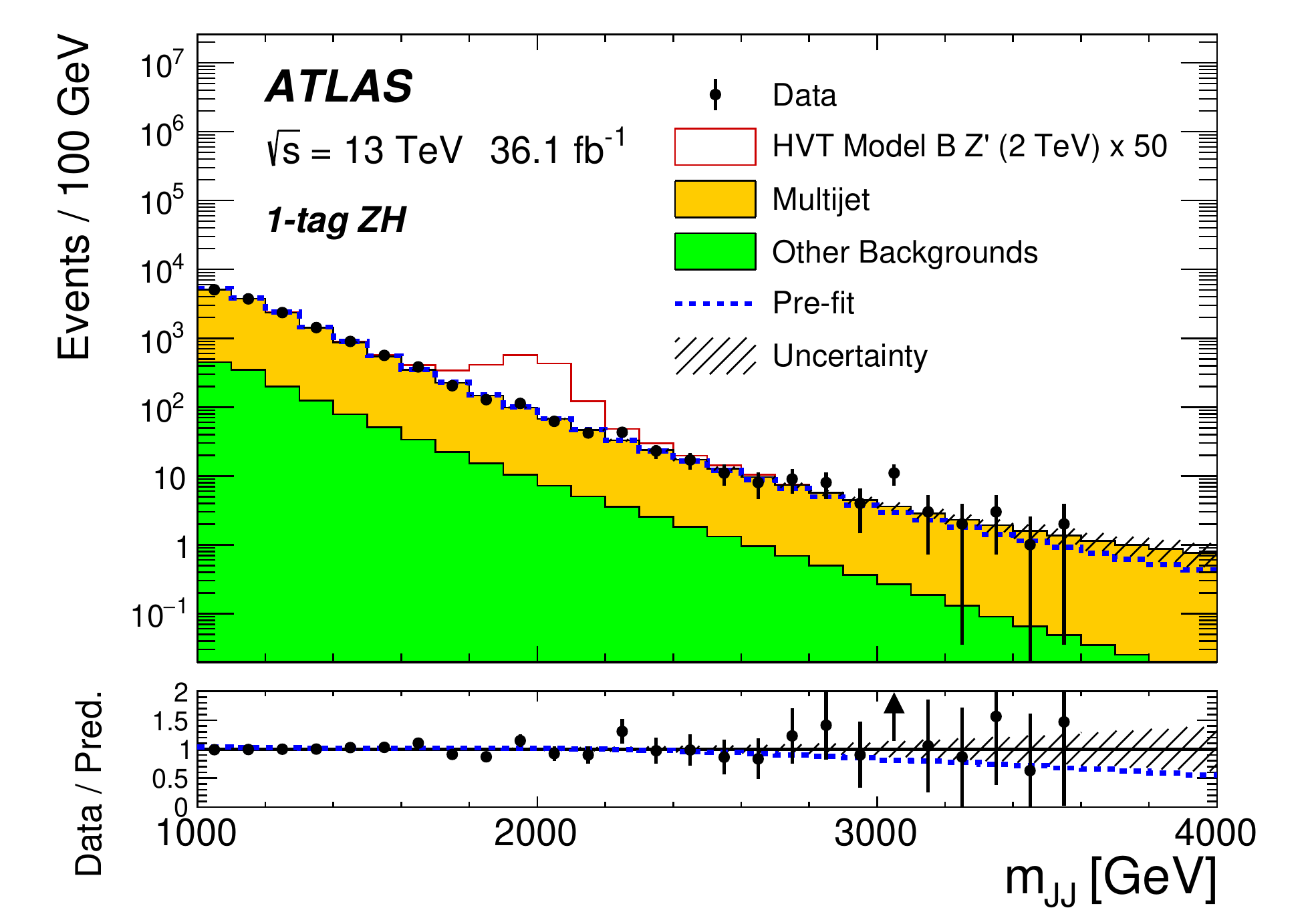}
\includegraphics[width=0.49\textwidth]{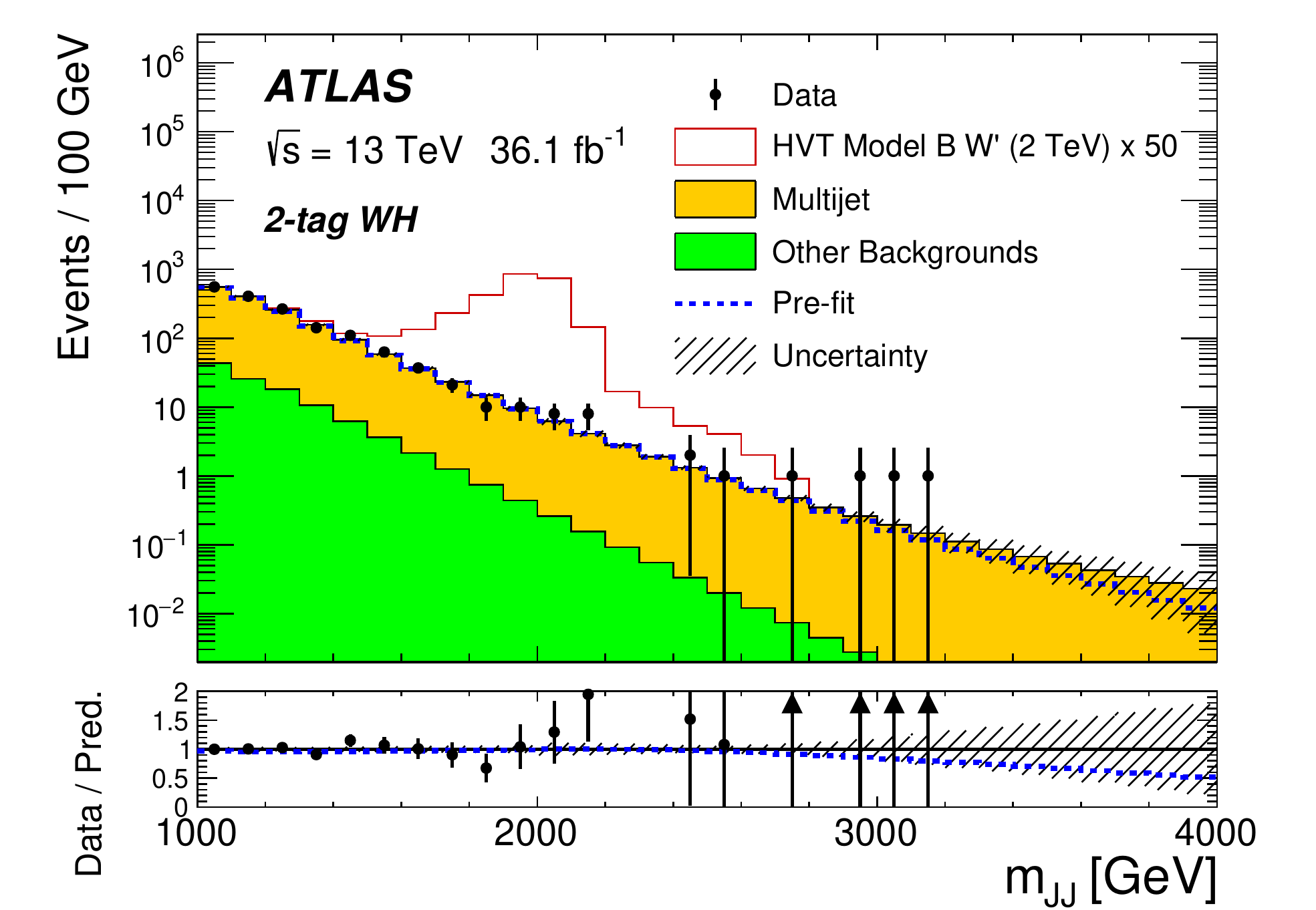}
\includegraphics[width=0.49\textwidth]{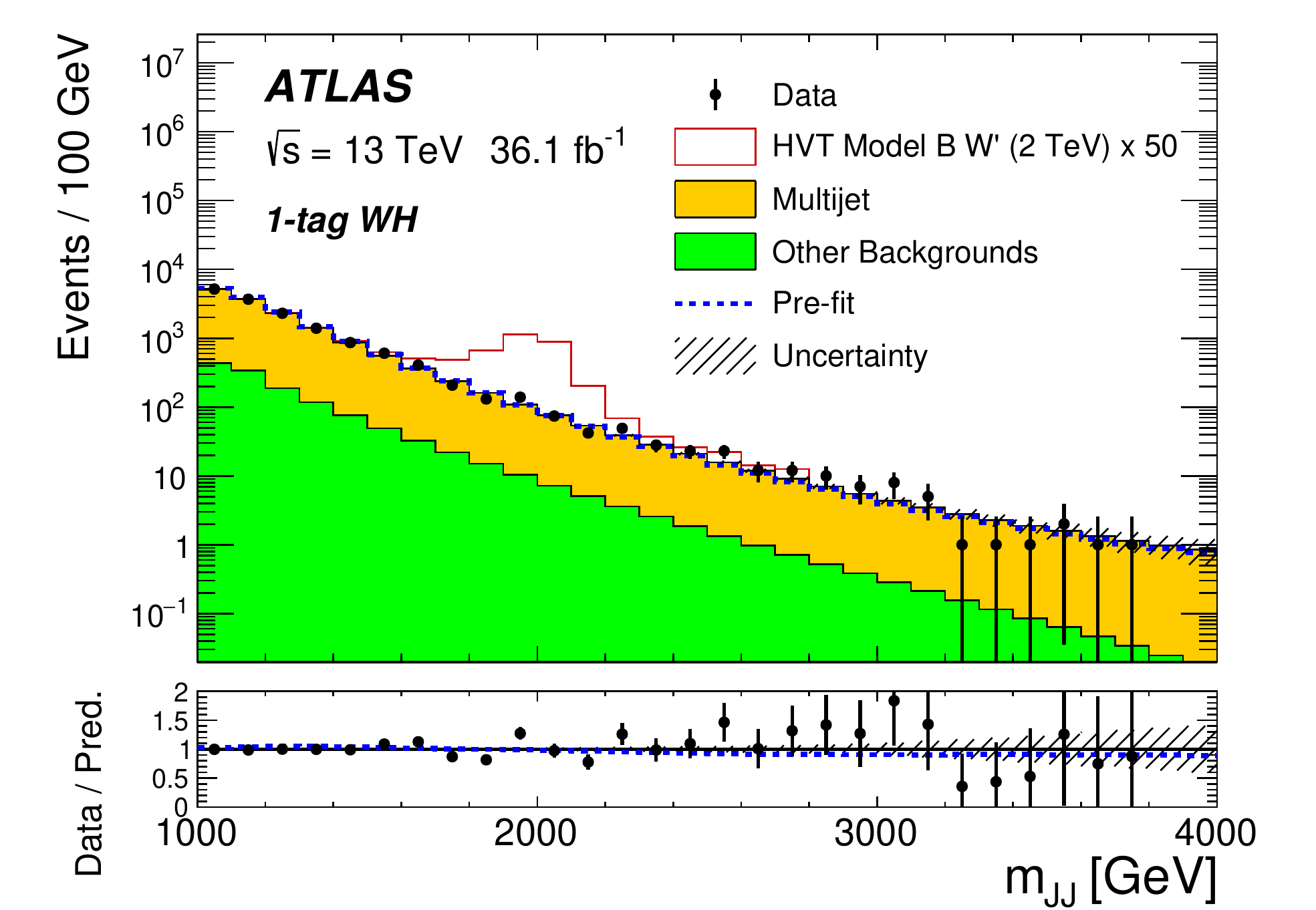}
\end{center}
\caption{The $m_{JJ}$ distributions in the $VH$ signal regions for data (points) and 
 background estimate (histograms) after the likelihood fit for events in the
 (left) 2-tag and (right) 1-tag categories. The pre-fit 
 background expectation is given by the blue dashed line.
 The expected signal
 distributions (multiplied by 50) for a HVT benchmark {\em Model B} $V^\prime$ boson with 2 \tev\ mass are also shown. 
In the data/prediction ratio plots, arrows indicate off-scale points.
}
\label{fig:m2J_SR}
\end{figure}

The number of background events in the 1-tag and 2-tag $ZH$ and $WH$ signal regions after the fit,
the number of events observed in the data, and the predicted yield for a potential signal
are reported in Table~\ref{tab:results}.  The total data and background yields in each region are constrained to agree by the fit.  There is a $\sim60\%$ overlap of data between the $WH$ and $ZH$ selections for both the 2-tag and 1-tag signal regions, and this fraction is approximately constant as a function of $m_{JJ}$.

\begin{table}[!htb]
\begin{center}
\caption{The number of predicted background
  events in the $VH$ 1-tag and 2-tag signal regions after the fit, compared to the data.  The ``Other backgrounds'' entries include both $t\bar{t}$ and $V$+jets.
  Uncertainties correspond to the total uncertainties in the predicted event yields, and are smaller for the total than for the individual 
  contributions because the latter are anti-correlated.
  The yields for $m = $ 2 \tev\ $V^\prime$ bosons decaying to $VH$ in {\em Model B} are also given. 
  Due to rounding the totals can differ from the sums of components.}
\begin{tabular}{lD{,}{\,\,}{-1}D{,}{\,\,}{-1}}
\toprule
 & \multicolumn{1}{c}{$ZH$ 2-tag} & \multicolumn{1}{c}{$ZH$ 1-tag} \\
\midrule
Multijet & 1440, \pm 60 & 13770, \pm 310 \\
Other backgrounds & 135, \pm 45  & 1350, \pm 270 \\
\midrule
Total backgrounds &  1575, \pm 40 & 15120, \pm 130 \\ 
Data &  1574,  & 15112, \\
\midrule
{\em Model B}, $m$ = 2 \tev & 25, \pm 7  & 29, \pm 10 \\
\toprule
 & \multicolumn{1}{c}{$WH$ 2-tag} & \multicolumn{1}{c}{$WH$ 1-tag} \\
\midrule
Multijet & 1525, \pm 65  & 13900, \pm 290 \\
Other backgrounds & 110, \pm 45  &  1310, \pm 260\\
\midrule
Total backgrounds & 1635, \pm 40  & 15220, \pm 120  \\
Data & 1646,  & 15212, \\
\midrule
{\em Model B}, $m$ = 2 \tev & 51, \pm 10  & 62, \pm 16 \\
\bottomrule
\end{tabular}
\label{tab:results} 
\end{center}
\end{table}

\subsection{Statistical analysis}
\label{subsec:stats}

To determine if there are any statistically significant local excesses in the data,
a test of the background-only hypothesis ($\mu=0$) is performed at each signal mass point.
The significance of an excess is quantified using the local $p_{0}$ value, the probability that the
background could produce a fluctuation greater than or equal to the excess observed in data.
A global $p_0$ is also calculated for the most significant discrepancy, using background-only
pseudo-experiments to derive a correction for the look-elsewhere effect across the mass range tested~\cite{Gross:2010qma}.
The largest deviation from the background-only hypothesis is in the $ZH$ signal region, occurring at $m_{JJ} \approx 3.0$~\tev\ with 
a local significance of \localSig~$\sigma$.
The global significance of this excess is \globalSig~$\sigma$.

The data are used to set upper limits on the cross-sections for the different benchmark signal processes.
Exclusion limits  are computed using the \cls method~\cite{Read:2002hq}, with a value of $\mu$
regarded as excluded at the 95\% CL when \cls~is less than 5\%.  

Figure~\ref{fig:CombinedLimits} shows the 95\% CL cross-section upper limits on HVT resonances for both \textit{Model A} and \textit{Model B}  in the $WH$ and $ZH$ signal regions for masses between 1.1 and 3.8~\tev.   Limits on \\ \mbox{$\sigma(pp\rightarrow V^{\prime} \to VH) \times B(H \to (b\bar{b}+c\bar{c}))$}\footnote{The signal samples contain Higgs boson decays to $b\bar{b}$ and $c\bar{c}$, but due to the branching ratios and $b$-tagging requirements the sensitivity is dominated by $H \rightarrow b\bar{b}$.} are set in the range of \WHlimitOneTev~fb to \WHlimitThreePointEightTev~fb and \ZHlimitOneTev~fb to \ZHlimitThreePointEightTev~fb in the $WH$ and $ZH$ signal regions, respectively.   These cross-section limits are translated into excluded \textit{Model B} signal mass ranges of 1.10 -- 2.50~\tev\  for $WH$ resonances and 1.10 -- 2.60~\tev\ for $ZH$ resonances.  The corresponding excluded mass ranges for \textit{Model A} are 1.10 -- 2.40~\tev\  for $WH$ resonances, and 1.10 -- 1.48~\tev\ and 1.70 -- 2.35~\tev\ for $ZH$ resonances.

\begin{figure}[!ht]
\begin{center}
\includegraphics[width=0.48\textwidth]{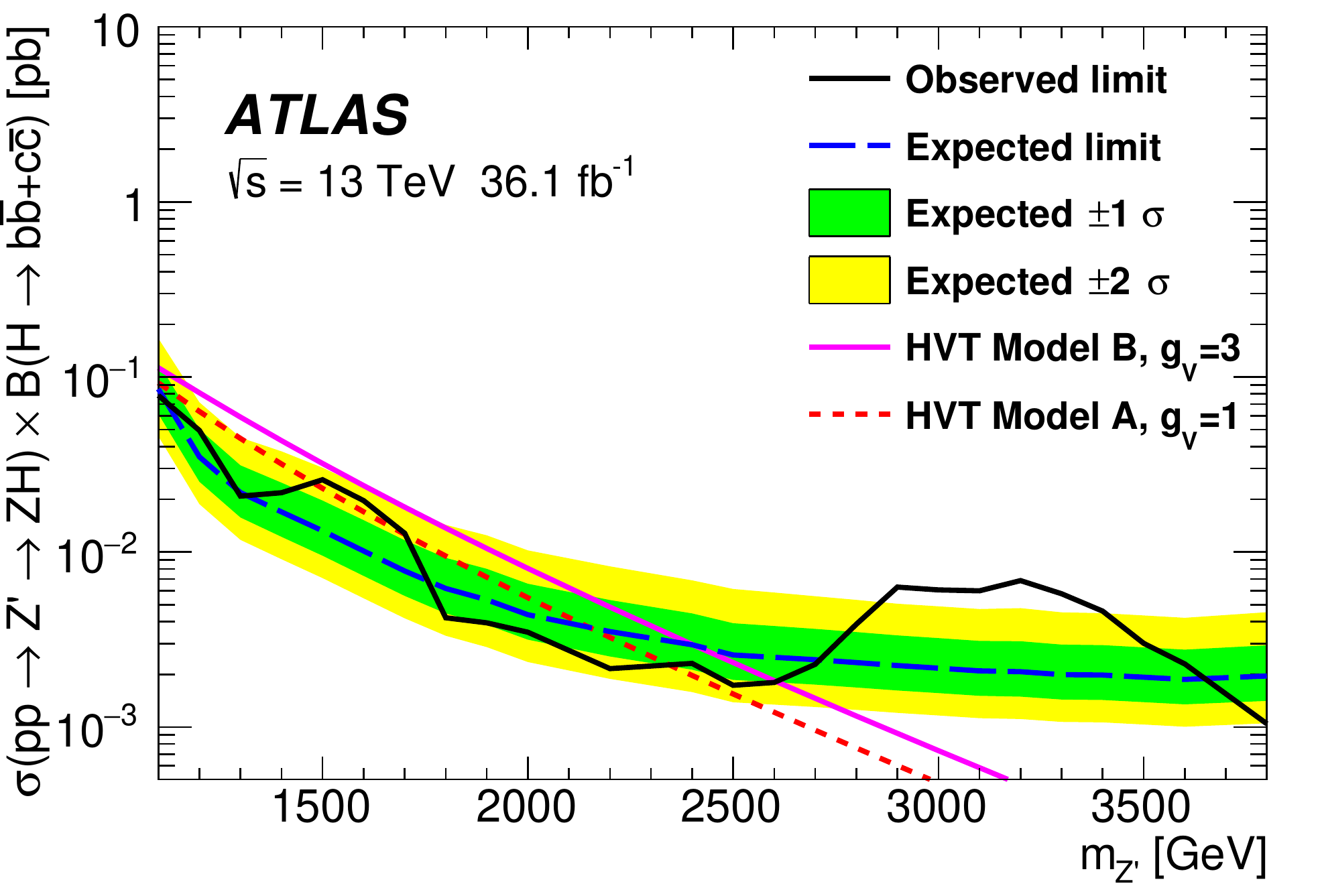}
\includegraphics[width=0.48\textwidth]{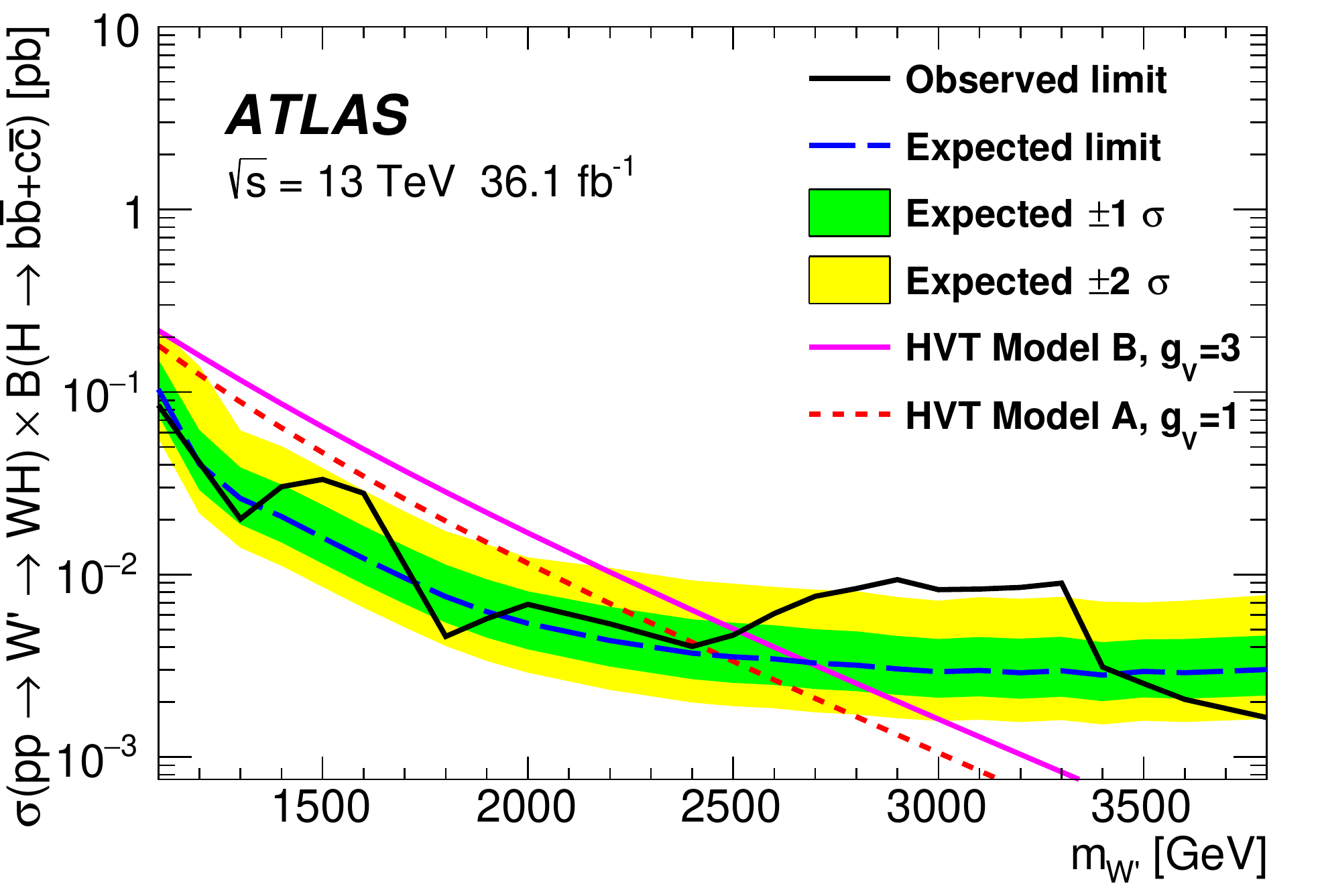}
\end{center}
\caption{The observed and expected cross-section upper limits at the 95\% confidence level for $\sigma(pp\rightarrow V^\prime \rightarrow VH) \times {\textrm B} (H \rightarrow (b\bar{b}+c\bar{c}))$, assuming SM branching ratios, in \textit{Model A} and \textit{Model B} in the (left) $ZH$ 
  and (right) $WH$ signal regions.
The red and magenta curves show the predicted cross-sections as a function of resonance mass for the models considered.
}
\label{fig:CombinedLimits}
\end{figure}

Figure~\ref{fig:lim2d} shows the 95\% CL limits in the $g^2 c_F / g_V$ vs. $g_V c_H$ plane for several resonance masses for both the $WH$ and $ZH$ channels. These limits are derived by rescaling the signal cross-sections to the values predicted for each point in the $(g^2 c_F / g_V, g_V c_H)$ plane and comparing with the observed cross-section upper limit. As the resonance width is not altered in this rescaling, areas for which the resonance width $\Gamma/m > 5\%$ are shown in grey.  These may not be well described by the narrow width approximation assumed in the rescaling.
\begin{figure}[htbp!]
\begin{center}
\includegraphics[width=0.48\textwidth]{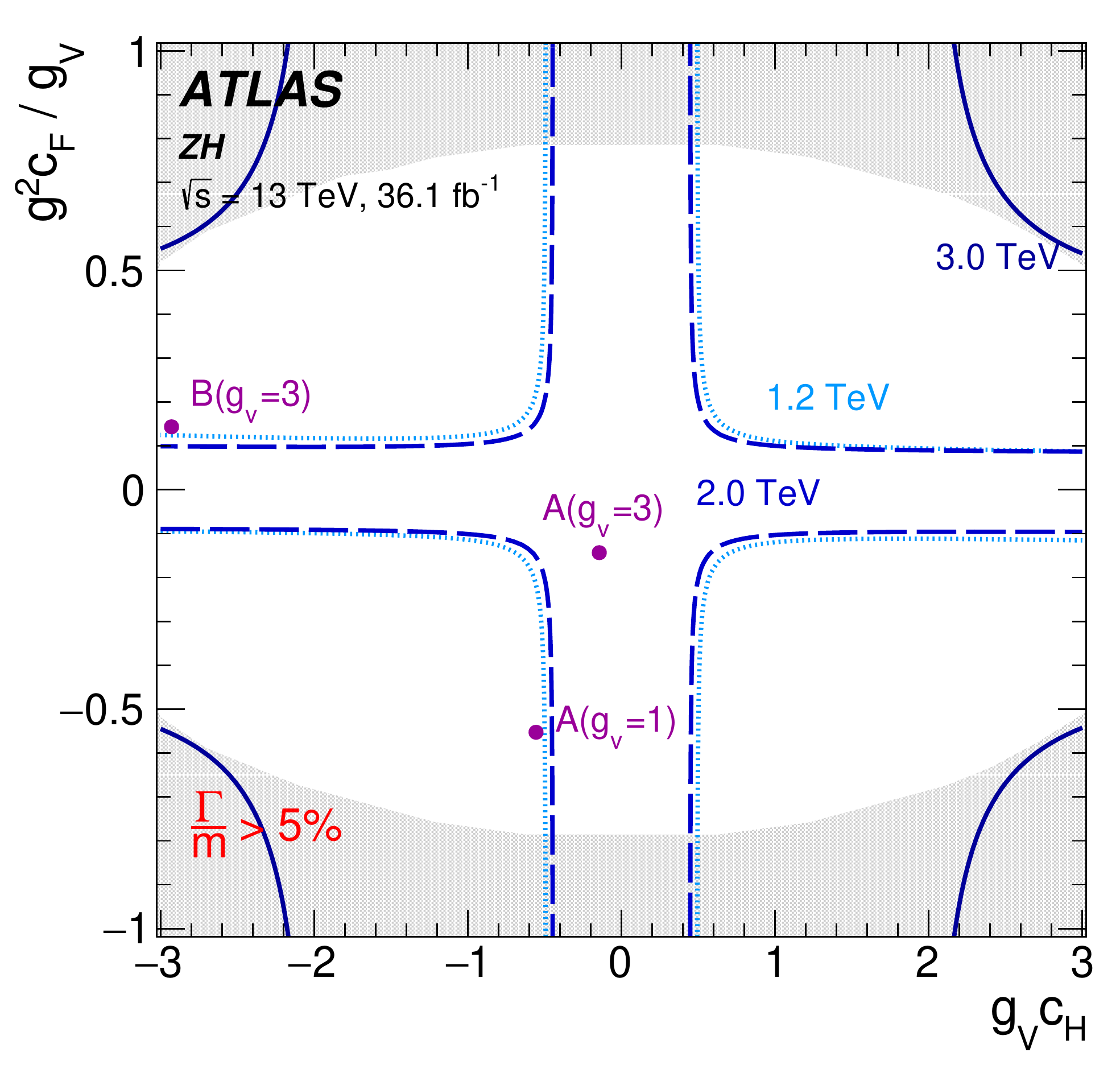}
\includegraphics[width=0.48\textwidth]{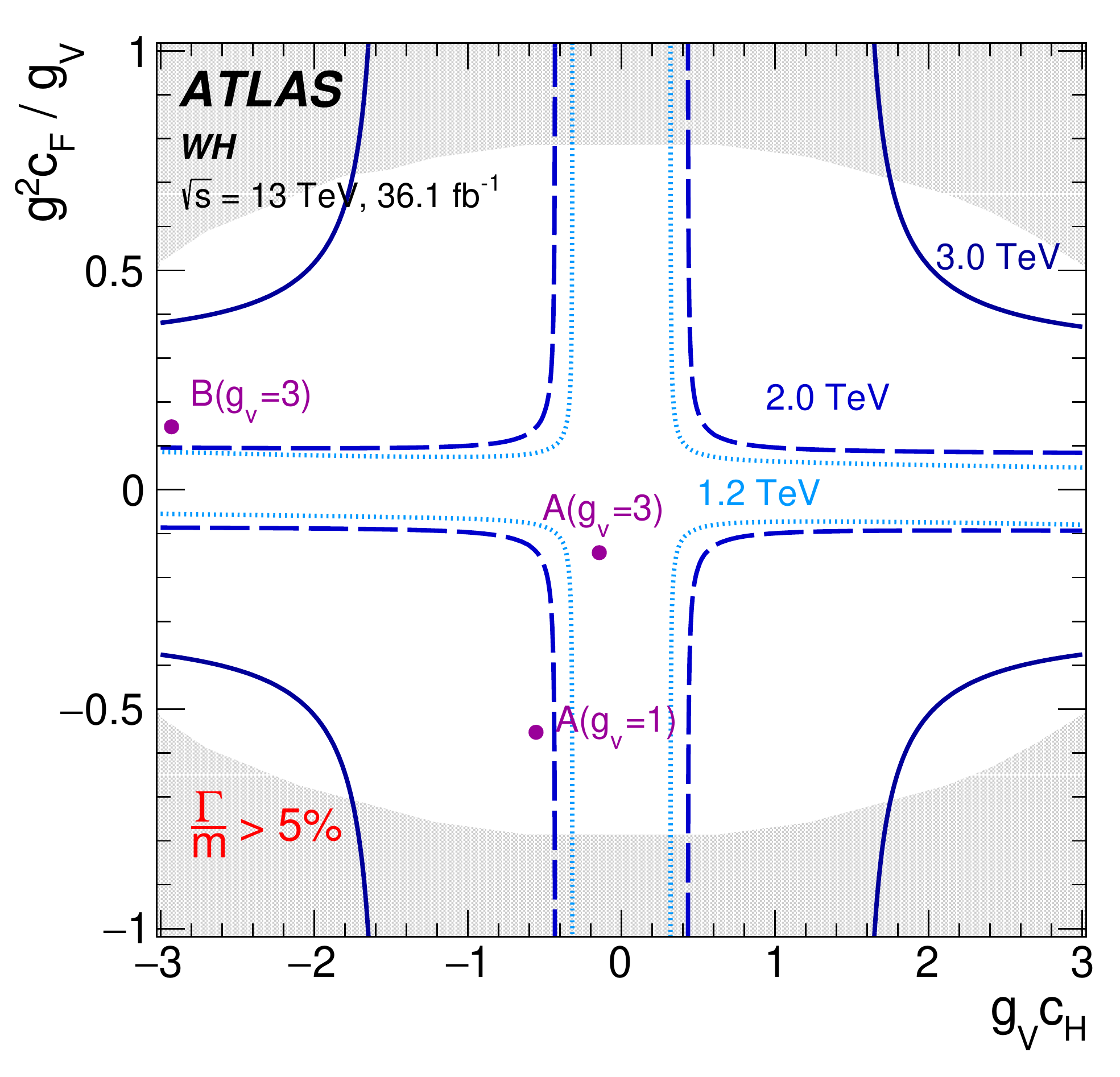}
\end{center}
\caption{Limits in the $g^2 c_F / g_V$ vs. $g_V c_H$ plane for several resonance masses for the (left) $ZH$ and (right) $WH$ channels.  Areas outside the curves are excluded.  The benchmark model points are also shown.  Coupling values for which the resonance width $\Gamma/m > 5\%$ are shown in grey, as these regions may not be well described by the narrow width approximation.
}
\label{fig:lim2d}
\end{figure}


\section{Summary}
\label{sec:conclusion}

A search for resonances decaying to a $W$ or $Z$ boson and a Higgs boson  has been
carried out in the $q\bar{q}^{(\prime)} b\bar{b}$ channel with \lumi\ fb$^{-1}$ of
$pp$ collision data collected by ATLAS during the 2015 and 2016 runs of the LHC
at $\rts = 13$ \tev.  Both the vector boson and Higgs boson candidates are reconstructed
using large radius jets, and jet mass and substructure observables are used to 
tag $W$, $Z$ and Higgs boson candidates and suppress 
the dominant multijet background.
In addition, small radius $b$-tagged track jets ghost-associated to the \largeR\ jets are 
exploited to select the Higgs boson candidate jet.
The data are in agreement with the Standard Model expectations, with the
largest excess observed at $m_{\textrm JJ} \approx 3.0$ \tev\ in the $ZH$ channel with a local 
significance of \localSig~$\sigma$.  The global significance of this excess is \globalSig~$\sigma$.
Upper limits on the production cross-section times the Higgs boson branching ratio to the
$b\bar{b}$ final state are set 
for resonance masses in the range between 1.1 and 3.8 \tev\
with values ranging from \WHlimitOneTev~fb to \WHlimitThreePointEightTev~fb and \ZHlimitOneTev~fb to \ZHlimitThreePointEightTev~fb (at 95\% CL)
for $WH$ and $ZH$ resonances, respectively.
The corresponding excluded heavy vector triplet \textit{Model B} signal mass ranges are 1.1 -- 2.5 \tev\  for $WH$ resonances, 
and 1.1 -- 2.6 \tev\ for $ZH$ resonances.

\section*{Acknowledgements}


We thank CERN for the very successful operation of the LHC, as well as the
support staff from our institutions without whom ATLAS could not be
operated efficiently.

We acknowledge the support of ANPCyT, Argentina; YerPhI, Armenia; ARC, Australia; BMWFW and FWF, Austria; ANAS, Azerbaijan; SSTC, Belarus; CNPq and FAPESP, Brazil; NSERC, NRC and CFI, Canada; CERN; CONICYT, Chile; CAS, MOST and NSFC, China; COLCIENCIAS, Colombia; MSMT CR, MPO CR and VSC CR, Czech Republic; DNRF and DNSRC, Denmark; IN2P3-CNRS, CEA-DSM/IRFU, France; SRNSF, Georgia; BMBF, HGF, and MPG, Germany; GSRT, Greece; RGC, Hong Kong SAR, China; ISF, I-CORE and Benoziyo Center, Israel; INFN, Italy; MEXT and JSPS, Japan; CNRST, Morocco; NWO, Netherlands; RCN, Norway; MNiSW and NCN, Poland; FCT, Portugal; MNE/IFA, Romania; MES of Russia and NRC KI, Russian Federation; JINR; MESTD, Serbia; MSSR, Slovakia; ARRS and MIZ\v{S}, Slovenia; DST/NRF, South Africa; MINECO, Spain; SRC and Wallenberg Foundation, Sweden; SERI, SNSF and Cantons of Bern and Geneva, Switzerland; MOST, Taiwan; TAEK, Turkey; STFC, United Kingdom; DOE and NSF, United States of America. In addition, individual groups and members have received support from BCKDF, the Canada Council, CANARIE, CRC, Compute Canada, FQRNT, and the Ontario Innovation Trust, Canada; EPLANET, ERC, ERDF, FP7, Horizon 2020 and Marie Sk{\l}odowska-Curie Actions, European Union; Investissements d'Avenir Labex and Idex, ANR, R{\'e}gion Auvergne and Fondation Partager le Savoir, France; DFG and AvH Foundation, Germany; Herakleitos, Thales and Aristeia programmes co-financed by EU-ESF and the Greek NSRF; BSF, GIF and Minerva, Israel; BRF, Norway; CERCA Programme Generalitat de Catalunya, Generalitat Valenciana, Spain; the Royal Society and Leverhulme Trust, United Kingdom.

The crucial computing support from all WLCG partners is acknowledged gratefully, in particular from CERN, the ATLAS Tier-1 facilities at TRIUMF (Canada), NDGF (Denmark, Norway, Sweden), CC-IN2P3 (France), KIT/GridKA (Germany), INFN-CNAF (Italy), NL-T1 (Netherlands), PIC (Spain), ASGC (Taiwan), RAL (UK) and BNL (USA), the Tier-2 facilities worldwide and large non-WLCG resource providers. Major contributors of computing resources are listed in Ref.~\cite{ATL-GEN-PUB-2016-002}.

%




\printbibliography

\newpage 
\begin{flushleft}
{\Large The ATLAS Collaboration}

\bigskip

M.~Aaboud$^\textrm{\scriptsize 137d}$,
G.~Aad$^\textrm{\scriptsize 88}$,
B.~Abbott$^\textrm{\scriptsize 115}$,
O.~Abdinov$^\textrm{\scriptsize 12}$$^{,*}$,
B.~Abeloos$^\textrm{\scriptsize 119}$,
S.H.~Abidi$^\textrm{\scriptsize 161}$,
O.S.~AbouZeid$^\textrm{\scriptsize 139}$,
N.L.~Abraham$^\textrm{\scriptsize 151}$,
H.~Abramowicz$^\textrm{\scriptsize 155}$,
H.~Abreu$^\textrm{\scriptsize 154}$
% ...
\bigskip
\\
% ...
$^{12}$ Institute of Physics, Azerbaijan Academy of Sciences, Baku, Azerbaijan\\
% ...
$^{88}$ CPPM, Aix-Marseille Universit{\'e} and CNRS/IN2P3, Marseille, France\\
% ...
$^{115}$ Homer L. Dodge Department of Physics and Astronomy, University of Oklahoma, Norman OK, United States of America\\
% ...
$^{119}$ LAL, Univ. Paris-Sud, CNRS/IN2P3, Universit{\'e} Paris-Saclay, Orsay, France\\
% ...
$^{137}$ $^{(a)}$ Facult{\'e} des Sciences Ain Chock, R{\'e}seau Universitaire de Physique des Hautes Energies - Universit{\'e} Hassan II, Casablanca; $^{(b)}$ Centre National de l'Energie des Sciences Techniques Nucleaires, Rabat; $^{(c)}$ Facult{\'e} des Sciences Semlalia, Universit{\'e} Cadi Ayyad, LPHEA-Marrakech; $^{(d)}$ Facult{\'e} des Sciences, Universit{\'e} Mohamed Premier and LPTPM, Oujda; $^{(e)}$ Facult{\'e} des sciences, Universit{\'e} Mohammed V, Rabat, Morocco\\
% ...
$^{139}$ Santa Cruz Institute for Particle Physics, University of California Santa Cruz, Santa Cruz CA, United States of America\\
% ...
$^{151}$ Department of Physics and Astronomy, University of Sussex, Brighton, United Kingdom\\
% ...
$^{154}$ Department of Physics, Technion: Israel Institute of Technology, Haifa, Israel\\
$^{155}$ Raymond and Beverly Sackler School of Physics and Astronomy, Tel Aviv University, Tel Aviv, Israel\\
% ...
$^{161}$ Department of Physics, University of Toronto, Toronto ON, Canada\\
% ...
$^{a}$ Also at Department of Physics, King's College London, London, United Kingdom\\
$^{b}$ Also at Institute of Physics, Azerbaijan Academy of Sciences, Baku, Azerbaijan\\
$^{c}$ Also at Novosibirsk State University, Novosibirsk, Russia\\
$^{d}$ Also at TRIUMF, Vancouver BC, Canada\\
$^{e}$ Also at Department of Physics {\&} Astronomy, University of Louisville, Louisville, KY, United States of America\\
% ...
$^{*}$ Deceased
\end{flushleft}



 
\end{document}